\documentstyle[12pt,aaspp4]{article}
\begin{document}

\title{Prospects for the Determination of Star Orbits Near the Galactic Center}

\author{M. Jaroszy\'{n}ski}
\affil{Warsaw University Observatory, Al.~Ujazdowskie~4,
00-478~Warszawa,Poland}
\affil{Visiting Scientists, Department of Astrophysical Sciences, Princeton
University}
\affil{e-mail: mj@sirius.astrouw.edu.pl}

\begin{abstract} 
We simulate the observations of proper motion of stars very close to the
Galactic Center. We show that the speckle interferometry done with the Keck
II telescope is accurate enough to obtain orbital parameters for stars
with the period $P \sim 10~{\rm y}$ during $\sim 10$ seasons of astrometric
observations made once a year. The determination of a single orbit will
give central mass estimate with the typical uncertainty of the existing
mass determinations based on velocity dispersion measurements.
A much higher precision orbits will be measured in several years when Keck
Interferometer becomes operational, and fainter stars are discovered even
closer to Sgr A*. Astrometry alone will provide accurate determination of $
M/D^3 $, where $ M $ is the black hole mass and $ D $ is the distance to
the Galactic Center.  If spectroscopic orbits of the stars are also measured 
then both: $ M $ and $ D $ will be precisely determined. 
\end{abstract}


\clearpage

\section{Introduction}

The proper motion studies of stars near the Galactic Center (Genzel et al.
1997; Eckart \& Genzel 1997; Ghez et al. 1998) show the astonishing
accuracy of the astrometric observations in the near infrared $K$ band. 
The closest to Sgr A$^*$ studied star is at the projected distance 
$\sim 100~{\rm mas}$
(corresponding to $\sim 850~{\rm AU}$ at $8.5~{\rm kpc}$)
from its position, and moves with the velocity $\sim 1400~{\rm
km/s}$ in the plane of the sky. The radial velocities of some of the stars
at $\ge 3\arcsec$ from Sgr A$^*$ are also measured (Genzel et al.
1996) and the comparison with the proper motion data shows that the
velocity distribution is nearly isotropic. The published observations cover
a relatively short span of time and the above authors use a statistical
approach to find the mass in the central $\sim 0.01~{\rm pc}$ region around
the Galactic Center. The present accuracy of the observations makes it
possible to study the orbits of individual stars and derive the mass
in the central part of the Galaxy by more direct methods, similar to those
used in classical binary systems of stars.

Recently Salim \& Gould (1998) proposed the study of the orbits of
individual stars in the vicinity of Sgr A$^*$ to get its distance. Such
measurement may be based on the accumulation of astrometric data augmented
by the radial velocity data. Salim \& Gould consider three stars with
the smallest projected distances from Sgr A$^*$ which can be found in 
the Ghez et al. (1998) catalog. They investigate the accuracy of the
Sgr A$^*$ distance estimate achieved after given observation
time and its dependence on the actual orbit inclinations and periods of the
chosen stars. They assume that the star positions will be obtained with the
present accuracy ($2~{\rm mas}$) and that the radial velocity will be
measured with an error smaller than $50~{\rm km/s}$.

In this paper we address similar questions using a different approach. 
First we are interested in the accuracy of the determination of all orbit
parameters and the accuracy of the determination of the central mass. We
are also interested in determinations based on better astrometric accuracy
and using fainter stars, which may in future be found closer to the
Galactic Center. We do not use any particular stars with already measured
proper motions, but rather simulate the orbits with given semimajor axis,
not exceeding $\sim 10^3~{\rm AU}$ and randomly chosen eccentricity and
orientation in space. Similar approach has been used by Jaroszy\'{n}ski
(1998b), in the study of the observability of relativistic effects in
motion of stars close to Sgr A$^*$. According to this paper only the
relativistic motion of periastron would be measurable for orbits 
$\sim 1000~{\rm AU}$ in size and only if the accuracy of astrometric
measurements is much higher than the present one, reaching the future
capabilities of the Keck Interferometer (van Belle \& Vasisht 1998). 
We neglect the relativistic effects altogether using purely Newtonian
star trajectories in our studies. 

While the presence of the black hole in the Galactic Center has not 
been proven yet, and the existence of a dense cluster 
of some kind of dark matter here (Munyaneza, Tsikaluri, \& Violler 1998) 
has not been excluded, we assume that there is in fact a point mass in the
Galactic Center. We adopt the central black hole mass estimate of Ghez et
al. (1998),  $M_0=2.6(\pm 0.2) \times 10^6~{\rm M}_{\sun}$,   and the
distance to the Galactic Center, $D_0=8.5~{\rm kpc}$ for our
simulations. Our analysis is aimed at finding the relative error in
possible mass and distance estimates and does not depend critically on their
exact values used for simulations. With the adopted distance and mass the
angular size of $100~{\rm mas}$ corresponds to $850~{\rm AU}$ and to the
orbital period of $15.4~{\rm y}$ for an elliptical orbit of this semimajor
axis. 

In the next Section we consider the modeling of orbits based on
astrometric observations alone. In Sec.~3 we consider fits based on the
combined astrometric and radial velocity data. In Sec.~4 we estimate the
number of stars in the close vicinity of Sgr A$^*$, which may in future be
used for black hole mass and its distance determination. The discussion
follows in the last Section.

\section{Simulations of Astrometric Observations of Star Motions}

We consider stars on elliptic orbits with ``true'' semimajor axis $a_0 \le
10^3~{\rm AU}$. The closest to Sgr A$^*$ star with measured proper motion
(Ghez et al. 1998) is at the projected distance $114~{\rm mas}$, 
so its 3D distance 
$r \ge 969~{\rm AU}$ and the semimajor axis of its orbit must be greater
than $485~{\rm AU}$. (It can be much greater, of course.) 
We are also interested in faint stars ($K \leq 17$, or fainter), 
which may in future be found at similar or smaller distances from 
the Galatic Center. We postpone the discussion of the probability of
finding such stars until Sec.~4. 

We assume the accuracy of the relative position measurements to be constant
in time and the observations to take place once a year, as has been the
practice until now.  
According to Ghez et al. (1998) the uncertainty of the relative position 
measurements for bright ($K \leq 15$) stars near Galactic Center
is typically $\sim 2~{\rm mas}$, which corresponds to $17~{\rm AU}$.
The proper velocity measurements are less accurate for fainter stars, 
with uncertainty doubling at each 2 magnitude interval. It suggests 
indirectly, that for faint stars ($K \leq 17$) the uncertainty 
in the relative position amounts to $\sim 4~{\rm mas}$, or $34~{\rm AU}$. 
These are the typical numbers we are using in our numerical experiments.
In the future the Keck Interferometer (van Belle and Vasisht 1998) will
achieve $\sim 20 {\rm \mu as}$ accuracy in astrometric mode, corresponding
to $0.17~{\rm AU}$. This number is another characteristic value,
which can be used for simulations.

We introduce a Cartesian coordinate system $(x_0,y_0)$ in the orbital 
plane with the origin at the position of the black hole and the
$x_0$ axis pointing toward the periastron. The ``true'' orbit is given
as: 
\begin{equation}
{2\pi \over P_0}(t-t_{0})=u-e_0\sin u;~~~
x_0=a_0(\cos u - e);~~~
y_0=b_0\sin u
\end{equation}
where $P_0$ is the orbital period, $t_{0}$ - time of the passage through the
periastron, $a_0$ and $b_0$ are the major and minor
semiaxes of the ellipse, $e_0$ is its eccentricity, and $u$ - eccentric
anomaly. The orientation of the ellipse in space is given by the
three angles (inclination $i_0$, position
angle  of the line of nodes in the sky $\Omega_0$, and the angle of the
periastron measured from the ascending node of the orbit $\omega_0$).
The position of the star in the sky is obtained after the
projection of its position in space, which is given as:
\begin{equation}
{\bf r}(t)=x_0(t){\bf e}_x+y_0(t){\bf e}_y
\end{equation}
where ${\bf e}_x$ and ${\bf e}_y$ are the 3D unit vectors along $x_0$ and
$y_0$ axes.

Our approach is a Monte Carlo simulation of synthetic data sets 
(e.g. Press et al. 1987). Usually one has a model fitted to the real data 
and is interested in the confidence limits on the estimated parameters. 
One of the possible way of doing it is to take the fitted parameters 
as true and simulate the sets of observations of the system assuming that
the model is a good representation of the system. In our case the 
known orbit parameters allow the calculation of the accurate star 
position on the sky at any time. The measured positions are in error. With
the estimated uncertainty in position measurements $\sigma$ we assume the 
measured position to be normally distributed:
\begin{equation}
p(X)dX={1 \over \sqrt{2\pi}\sigma}\exp{\left(-(X-X_0)^2/2\sigma^2\right)}dx
\end{equation}
where $X$ can be any of the two measured coordinates of the star on the sky,
and $X_0$ is its ``true'' value. We draw randomly the simulated positions 
of the star from the probability distribution and obtain the synthetic 
data set. Repeating the procedure we get many such sets.
Fitting  models to these simulated observations we obtain different sets of
model parameters scattered around the original values. The scatter in
these fitted parameters is a good measure of the confidence limits
of the original fit. Thus starting with an orbit of known parameters 
and simulating many synthetic data sets with the same uncertainty $\sigma$,
we can learn about a likely quality of the fitted model. In particular we 
can estimate the typical errors of the fit.

The mass of the black hole is related to the orbital parameters through the
Kepler's law: 
\begin{equation}
M={4\pi^2a^3 \over {\rm G}P^2}
\end{equation}
where $G$ is the gravitational constant.
In this Section we assume that the distance to the Galactic Center is known
and equal to $D_0=8.5~{\rm kpc}$, so the directly measurable angular sizes
are equivalent to corresponding linear sizes. In general the quantity which
can be estimated from the astrometric data alone is the ratio
$M/D^3$. 
In our simulations of star orbits we use $M_0=2.6 \times 10^6~M_{\sun}$ for
the value of the black hole mass. The mass estimated from 
the models fitted to simulated data is scaterred around $M_0$.
For different central mass values of some of the fitted parameters  would
scale, but the procedure would remain the same.

We find the model parameters using the least square minimization of
the expression:
\begin{equation}
\chi^2=\sum_{j=1}^{N}~
{({\bf X}_j-{\bf X}(t_j;a,e,P,i,\Omega,\omega,t_0))^2
\over \sigma^2}
\end{equation}
where ${\bf X}_j$ is the  ``measured'' position of the star
at the time $t_j$
and ${\bf X}(t_j;a,e,P,i,\Omega,\omega,t_0)$ is the position
resulting from a model with the given parameters and calculated for 
the same instant of time.

The semimajor axis $a_0$ of the ``true'' orbit serves as a main parameter 
of our simulations. Our study shows that the quality of the fits depends
mostly on this parameter (and on the ``true'' period, since the two are
related). For practical reasons we limit the number of iterations in the
procedure finding the minima of  $\chi^2$. A deeper analysis of the fitting
shows a weak dependence of its success on the orbit eccentricity, showing
that the cases with $e \approx 1$ are relatively more difficult. We neglect
this fact in our simulations, which means, that the orbits having the above
property are slightly underrepresented among successful fits. 
For a given $a_0$ we choose the eccentricity $0 \leq e_0 \leq 1$ 
as a random number. The $\cos i_0$, $\Omega_0$, and $\omega_0$ are also
given random values to guarantee the isotropic distribution of orbits
orientation and position of the periastron. The time of periastron passage
has no physical meaning (any properties of the motion depend on $t-t_0$
only) so we choose it at random from the range $0 - P$. 

We assume observations to take place in a randomly chosen day of June and
to be repeated through $N$ seasons. The basic number of observations we
consider is $N=10$,  but we also make fewer simulations for other numbers.
For each synthetic data sets $\{{\bf X}_j\}$ we find a model, starting the
fitting procedure from the ``true'' values of the parameters. Since we expect
the parameters fitted to the scattered data to be close to the ``true''
parameters, this starting point seems to be the best. If the fit converges,
and if the obtained minimal $\chi^2$ is smaller than the tabularized value
for given number of the degrees of freedom and required confidence, we include
the parameters of the model to the sample. Otherwise we neglect them, but
we keep the track of such unsuccessful fits.

The results of our simulations are shown in Figs 1,2. In the upper  panel
of Fig. 1 we show the ratio of the median value of the fitted semimajor
axis $a$ to its ``true'' value $a_0$. We also draw the lines showing the
region including 68\% of the sample points. Since the accuracy of position
measurement is constant,  the relative errors in fitted values of $a$ are
larger for smaller orbits. 
The opposite can be said of the accuracy of the period estimates, which become
more accurate when the total span of observations becomes longer than the
orbital  period. That means increasing accuracy for smaller orbits.
The eccentricity is related to the orbit shape and can be better
estimated for large orbits. In Fig.2 we show the result for mass estimation
based on the estimation of the orbital parameters. In this plot we see that
the estimates for small orbits become less accurate. Even if there are
faint stars on close orbits near the Galactic Center, the speckle
interferometry with the Keck telescope, with the position uncertainty of
$\sim 2~{\rm mas}$ ($17~{\rm AU}$) is not sufficient to give mass estimates
better than obtained with the  existing methods. For the stars on large
($\sim 10~{\rm y}$) orbits the ``once a year'' strategy seems promising,
but requires several years of data acquisition.

\section{Simulations of Astrometric and Radial Velocity Observations}

The measurements of the radial velocities for some of the Galactic Center
stars with measured proper motions have been done (Genzel et al. 1996) with
the accuracy of $\sigma_v=30~{\rm km/s}$. These stars are rather far from
the center (at projected distance $\ge 3^{\prime\prime}$), but similar
measurements for stars closer to the center, in the crowded field of view,
may be possible in the future. We optimistically assume that the same accuracy
of radial velocities will be possible for the stars closer to the center.  

With radial velocities measured and orbits determined from the proper
motion study, it is possible to estimate the distance $D$ to the source
(Salim \& Gould 1998). We use now $\alpha_0$ - the angular measure  of the
orbit semimajor axis as an independent parameter. (One has $a_0 \equiv
D\alpha_0$; $b_0 \equiv D\beta_0$.)

The velocity components of a star moving on an elliptic orbit are:
\begin{equation}
v_{0x}=-{2\pi D\alpha_0 \over P_0}~{\sin u \over 1-e\cos u};~~~
v_{0y}=+{2\pi D\beta_0 \over P_0}~{\cos u \over 1-e\cos u}
\end{equation}
where we use the reference frame of equation (1).
The velocity vector in space is
given as:
\begin{equation}
{\bf v}_0=v_{0x}(t){\bf e}_x+v_{0y}(t){\bf e}_y
\end{equation}
and its component along the line of sight can (in principle) be measured.
The parameter 
$V_0 \equiv 2\pi D\alpha_0/P_0$ measures the amplitude of the velocity and
can be independently fitted using the radial velocity data. The model of
the orbit including radial velocities has eight parameters and can be
fitted after the minimization of the expression
\begin{equation}
\chi^2=\sum_{j=1}^{N}~
{({\bf X}_j-{\bf X}(t_j;\alpha,e,P,i,\Omega,\omega,V,t_0))^2
\over \sigma^2}
+\sum_{j=1}^{N_v}~
{(v_j-v(t_j;\alpha,e,P,i,\Omega,\omega,V,t_0))^2
\over \sigma_v^2}
\end{equation}
where $N_v$ is the number of radial velocity measurements, $v_j$ is the
j-th measured radial velocity, and
$v(t_j;\alpha,e,P,i,\Omega,\omega,V,t_0)$ is 
the radial velocity at the instant $t_j$ resulting from the model with
given parameter values. 

With the velocity measured independently, the central mass and its distance
can be estimated: 
\begin{equation}
M={PV^3 \over 2\pi {\rm G}}
\end{equation}
\begin{equation}
D={VP \over 2\pi \alpha}
\end{equation}
where all the variables in the RHS are given by the fit. In Figure~3 we show
the results of our simulations including radial velocity measurements. The
results of the velocity fitting, mass estimate, and distance estimate are
shown. As can be seen in the plots, the relative error in velocity
fit becomes smaller for smaller orbits and the same can be said about the
mass estimation. Ten astrometric observations with accuracy of $\sim 2~{\rm
mas}$ with  5 radial velocity measurements done in the period of $\sim 10$
years are sufficient to give the distance to the Galaxy Center with a
accuracy better than 5\% ($1\sigma$) for large enough orbits  ($\ge
200~{\rm AU}$).  

\subsection{Accuracy of parameter fitting}

The improvement of interferometric equipment is expected to give much
better accuracy of position measurements, reaching $\sigma \sim 20~{\rm \mu
as}$ (van Belle \& Vasisht 1998) in the case of the Keck Interferometer. 
We investigate the
influence of the position accuracy on the expected errors in fitted
parameters, for the whole range of $\sigma$ from $20~{\rm \mu as}$ to 
$4~{\rm mas}$ ($0.17$ to $34~{\rm AU}$). We repeat our simulations for
several values of $\sigma$ and several values of the ellipse semimajor
axis, using the same ``observational strategy'' as above. 
``Observations'' of each orbit of a star are simulated many times. For
every simulation we get a set of estimated orbit parameters. 
We define the scatter in the estimated values of a parameter $p$ 
(where $p \in \{a,e,P,t_0,i,\Omega,\omega,V\}$) of a given
orbit as
\begin{equation}
\delta p =(p_{+} - p_{-})/2
\end{equation}
where 16\% of the estimated values of $p$ are greater than $p_{+}$, another
16\% are below $p_{-}$, and the remaining 68\% are between them. 
The scatter in the estimation of the parameter $p$  for all the orbits of
the same size $a_0$, observed with the same position accuracy $\sigma$ is
given as:
\begin{equation}
\Delta p = <\delta p>
\end{equation}
In Fig.4 we show the scatter in the fitted parameters. The ratio
$\sigma/a_0$ is a good estimate of the deformation introduced to the visual
orbit, so we use it as the abscissa for our plots. In the left column we
show the results for simulations based on astrometric measurements
alone. In the right column the typical errors in the orbit elements fitting
are shown for combined astrometric and radial velocity synthetic data.
It can be seen, that the radial velocity data, which has fixed accuracy in
our simulations, can improve the fitting procedure in case of poor
astrometric accuracy. 

We consider also the accuracy of mass and distance determinations. In
Fig.~5 we display the typical uncertainty in mass estimation based on two
methods and the results for the distance determination. Again, the
increased astrometric accuracy does not help much the determinations based
on radial velocity data.

\section{Possibility of Observing Stars Closer to the Center}

The central star cluster (Genzel et al. 1996, 1997) is the dominating stellar
component  within $\sim 10^2~{\rm pc}$ from the Galaxy Center. The density
distribution follows the softened isothermal sphere model 
\begin{equation}
n(r)={n_c \over 1+(r/r_c)^2}
\end{equation}
with the core radius $r_c=0.22~{\rm pc}$ ($5.33~{\rm arc sec}$). 
Projection along the line of sight gives the surface concentration of
stars:
\begin{equation}
{\cal N}(R) = \pi n {r_c^2 \over \sqrt{r_c^2+R^2}}
\equiv {\cal N}_K {r_c \over \sqrt{r_c^2+R^2}}
\end{equation}
where $R$ is the distance from the black hole measured in the plane of the
sky, and ${\cal N}_{17}=20~{\rm arcsec}^{-2}$ for stars brighter than
$K=17^m$. In the whole region of interest to us ($R \ll r_c$) the surface
density of stars belonging to the cluster core is constant.

According to Ghez et al. (1998) the sample of stars in their proper motion
studies constitutes a distinct cluster  with the core radius 
$r_{c1} = 0{\farcs}3$ ($\sim 0.01~{\rm pc}$) and the peak surface
density of  ${\cal N} \approx 15~{\rm arcsec}^{-2}$. Since the central
surface densities of both clusters are similar the volume density in the
smaller one is $\sim r_c/r_{c1}$ times larger than the volume density in
the core of the background cluster. Thus approximately half of the stars
that could be seen very close to Sgr A$^*$ on the sky are indeed the stars
very close to the center in 3D. 

The proper motion sample of stars (Ghez et al. 1998) is not complete and
probably cannot be used as a tracer of the general population of stars in
the very center of the Galaxy. On the other hand we expect that the future
proper motion studies of the Galaxy Center will employ similar selection
criteria, so the resulting samples will have similar space distribution. 

We are interested in the density of the
observable stars which are well inside the core. The observability of
sources is limited by the spatial resolution of the interferometer at given
limiting brightness. 
For the spatial resolution $d$ of the interferometer the density of stars
should not be too big:
\begin{equation}
{\cal N}(2d)^2 \le 1
~~~~\Rightarrow~~~~{\cal N}_{\rm max} \sim {1 \over 4d^2} 
\sim 10^4~{\rm arcsec}^{-2}
\end{equation}
where we adopt $d \approx 5~{\rm mas}$ as the resolution of the Keck
Interferometer. 

The combined surface star density in both clusters for $K \le 17^m$ is
${\cal N}_{17} \approx 35~{\rm arcsec}^{-2}$ (Genzel et al. 1997; Alexander
\& Sternberg 1998; Ghez et al. 1998). The maximal surface density of stars
${\cal N}_{\rm max}$ is $\sim 300$ times larger. The integral
luminosity function for the Galaxy Center has the slope $\beta=0.875$ at
$K=17^m$ (Blum et al. 1996). This slope flattens for stars less massive
than $\sim 0.7~M_{\sun}$ (Holtzman et al. 1998), which corresponds to 
$K \approx 21^m$ (Alexander \& Sternberg 1998) if we adopt the extinction
$A_K=3.5$ (Blum et al. 1996) to the Galactic Center. Rescaling to $21^m$ in
$K$ we have  ${\cal N}_{21} \approx 25~{\cal N}_{17}$, much less than the
maximal surface density introduced above. Thus the possibility of finding
faint stars close to the Galactic Center is limited by their volume density
and not by the limited resolution of the Keck Interferometer.

Using the same luminosity function to the proper motion sample of Ghez et
al. (1998) we find that there should be $\sim 25$ times more stars with
$K \le 21$ and with measurable proper motions. Some of them may be located
closer to the black hole than the stars already observed. For such faint
stars the Keck Interferometer operating in the imaging mode will have
position accuracy of $\sim 3~{\rm mas}$. The highest accuracy for
astrometry ($\sim 20 {\rm \mu as}$) will be possible for stars much
brighter, $K \le 17.6$ (van Belle \& Vasisht 1998). 

\section{Discussion and Conclusions}

We have investigated elliptical orbits of stars in the Newtonian potential
of a point mass. According to Jaroszy\'{n}ski (1998b)  the periastron
motion of the orbit due to the relativistic effects may be measurable for
small enough orbits ($a \lesssim 10^3~{\rm AU}$). Similar effect, but of
different sign, may be caused by the significant amount of matter
distributed continuously around the central black hole. The gravitational
lensing in the vicinity of Sgr A$^*$ (Jaroszy\'{n}ski 1998a) may deform a
part of the visual orbit if the observer is close to the orbital plane and
when the star is behind the central mass. All these effects are easy to
account for and can be introduced to the model. We neglect them here, since 
they have no significant influence on the accuracy of parameters fitting or 
mass and distance estimates.

Since we limit the number of iterations in the procedure finding $\chi^2$
minima, we have investigated the influence of this fact on the estimated
errors in fitted parameters. We are interested mostly in the semimajor
axis, period and mass estimates. As our analysis shows, the number of
iterations necessary for the fitting procedure to converge increases with
increasing orbit eccentricity. We have performed extra calculations
for orbits with the true semimajor axis $a_0=800{\rm AU}$, eccentricity
changing from $e_0=0.01$ to $0.99$ and random orientation in space,
simulating the astrometric observations of the moving stars with the
position accuracy of $\sigma=17{\rm AU}$.
The calculations show that the convergence is reached in $\sim 98$\% of 
the cases for $e_0 \le 0.8$ and falls to $\sim 90$\% for $e_0=0.99$.  
The quality of the fits, as measured by the value of $\chi^2$ does not
depend on the orbit eccentricity and the same is true about the scatter in
the estimated orbital period. The relative error in the estimated semimajor
axis is up to $\sim 2$ times larger for highly eccentric orbits 
($e_0 \ge 0.8$) as compared to lower eccentricity orbits. Since this group of
the orbits is underrepresented, the error is underestimated, but
only slightly. (An analysis taking into account the adequate number of high
eccentricity orbits would give $\lesssim 1.01$ times larger error estimates
for semimajor axis and $\lesssim 1.03$ times larger scatter in mass
estimates. Calculations with doubled number of allowed iterations confirm
this reasoning.) Similar investigation shows, that the orbit inclination
has negligible influence on the accuracy of the estimated semimajor axis,
period, or central mass.

The investigation of the proper motion of stars at distances $\lesssim
10^3~{\rm AU}$ from Sgr A$^*$ can provide a robust test of the existence of
a black hole there. If the mass is indeed in the form of a black hole, 
and the amount of mass distributed continuously is insubstantial, than
the stars should move on elliptical orbits and the rate of periastron
motion should agree with the mass estimated from the orbit size and
period. Each orbit directly probes the distribution of mass at distances 
$(1-e)a \le r \le (1+e)a$. Knowing few orbits one may cover a substantial
range in distances from the central mass. For such a test faint stars,
which may be found closer to the center than already observed relatively
bright stars with $K \leq 17$, can be used. The stars closer to the center
have shorter periods, so their orbits can be found in shorter time. The
increased accuracy of the astrometric position measurements will give very
accurate orbits for bright stars. This may eventually serve as a test of
point mass hypothesis at distances $\gtrsim 10^3~{\rm AU}$.

The measurement of radial velocities allows for the measurement of the
distance to Sgr A$^*$ (Salim \& Gould,998) and the absolute estimate of its
mass. With the present accuracy of astrometry ($2~{\rm mas}$) and
spectroscopy ($30~{\rm km/s}$) in $K$ band, and with ``observational
strategy'' adopted in our study, the most promising for the  distance
estimate are the orbits with $a \approx 600~{\rm AU}$, which would give
$\sim 3$\% accuracy of the measurement in 10 years. Better accuracy and 
for slightly larger orbits can be obtained after longer time (Salim 
\& Gould 1998). The mass estimate based on the radial velocity becomes 
more accurate for smaller orbits. Their investigation is so challenging 
observationally, that it is probably better to use large orbits and wait 
longer for the results.

\acknowledgements{Special thanks are due to Bohdan Paczy\'{n}ski for
suggesting the topic of this research, interest in its progress and kind
hospitality during my stay at the Department of Astrophysical Sciences,
Princeton University. This project was supported with the  NASA grant
NAG5-7016 to B.\ Paczy\'{n}ski and the Polish KBN grant 2P03D-012-12 to M.\
Jaroszy\'{n}ski. 
}

\clearpage

\figcaption[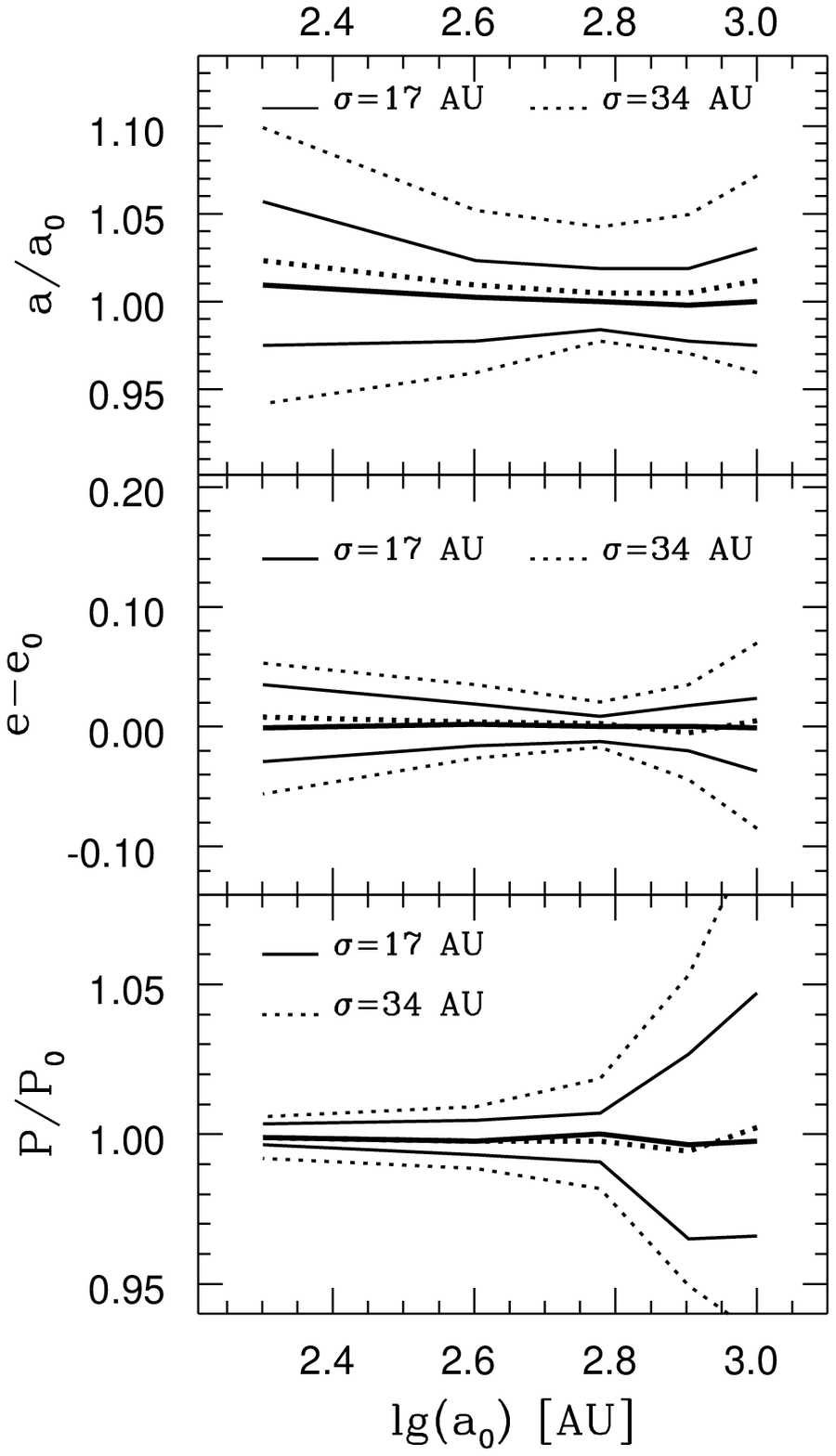]{
The median values of the ratio of the fitted orbit semimajor axis to the 
``true'' value of this parameter $a/a_0$ (top panel), 
the median value of the difference $e-e_0$ (middle panel), and the ratio of
the fitted period to its ``true'' value $P/P_0$ (lower panel) are drawn as
functions of the ``true'' value of the semimajor axis $a_0$ with heavy
lines. The thin lines encompass the region containing 68\% of the fitted
parameters values. 
Two cases ($\sigma=17~{\rm AU}$ and $\sigma=34~{\rm AU}$) are plotted.
The results are obtained for $N=10$ ``observations'' taken once a year.
The semimajor axis of 1000 AU corresponds to 118 mas in the sky
for the galactocentric distance of 8.5 kpc.
}

\figcaption[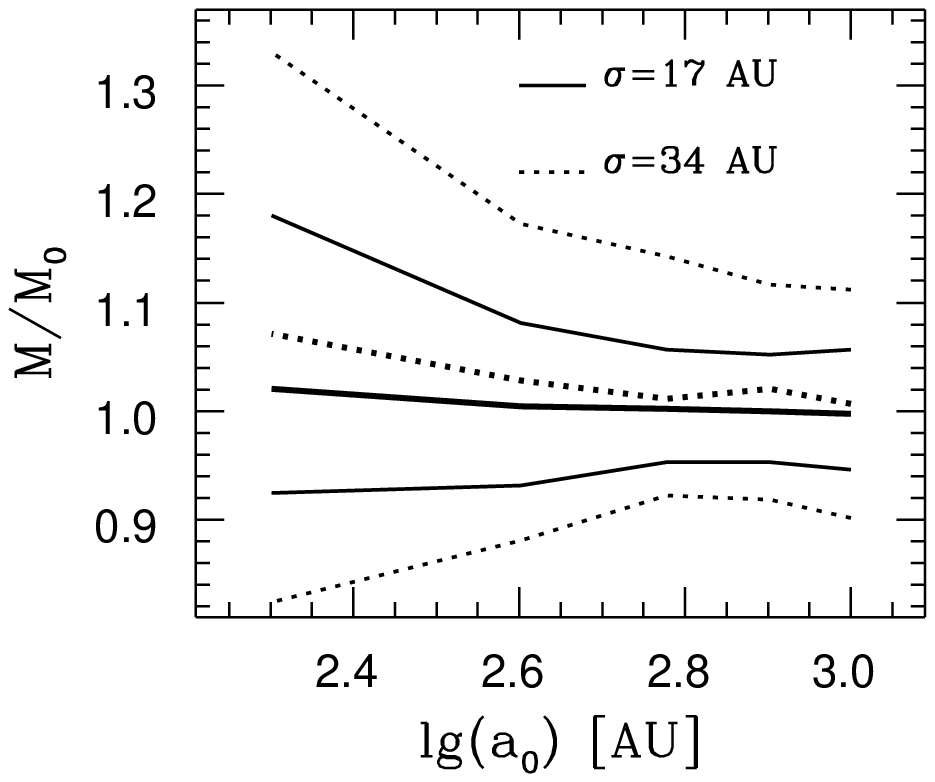]{
The ratio of the fitted mass to the black hole mass used in simulations,
$M/M_0$, shown as a function of the ``true'' semimajor axis $a_0$. The
distance to the Galactic Center is assumed to be $D_0=8.5~{\rm kpc}$. The
conventions follow Fig.~1, and the number of ``observations'' is $N=10$.
The semimajor axis of 1000 AU corresponds to 118 mas in the sky 
for the galactocentric distance of 8.5 kpc. 
}

\figcaption[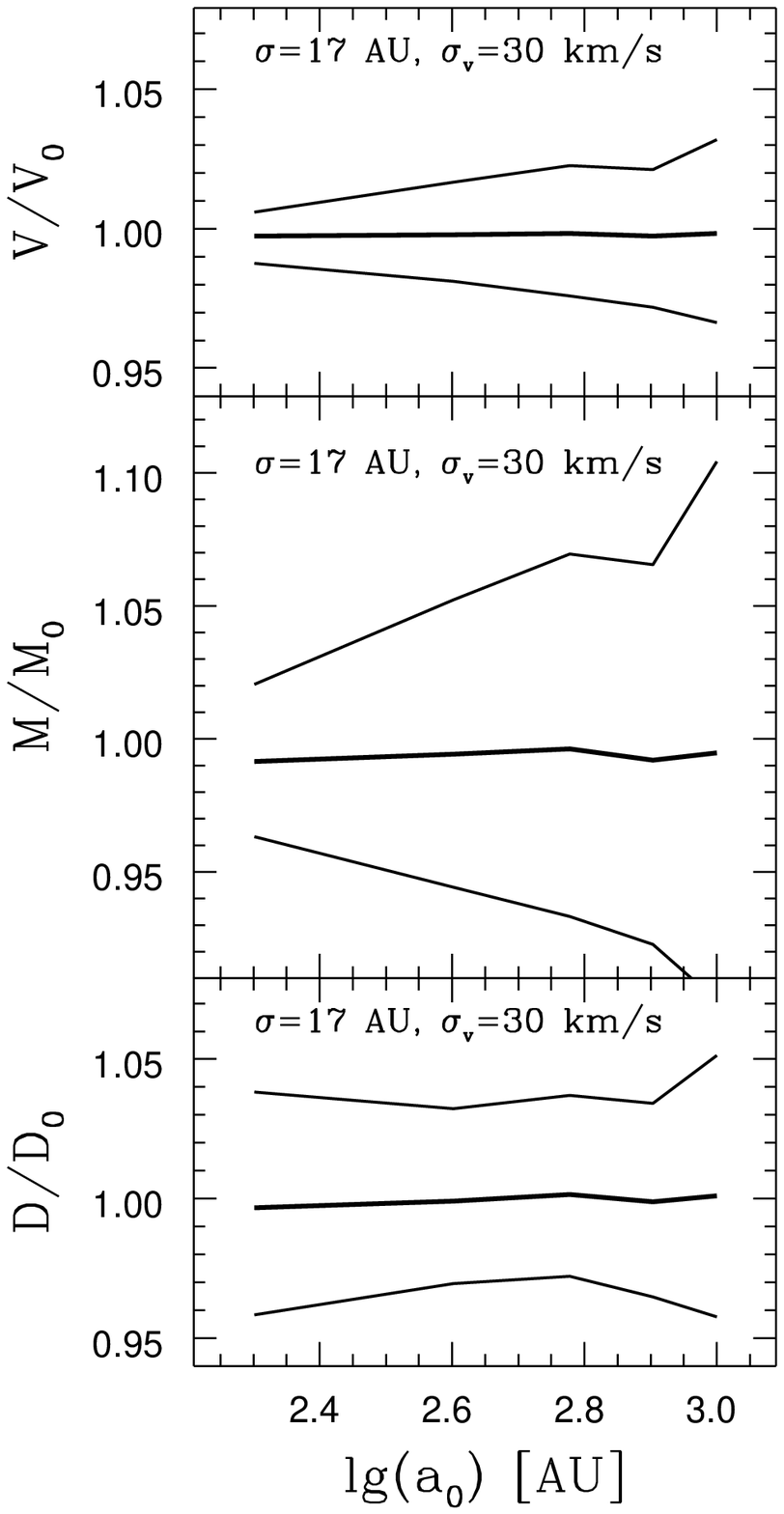]{
The ratios of the fitted velocity (upper panel), mass (middle), and distance
(lower) to the corresponding ``true'' values used in simulations. The mass
estimate is based on the fitted velocity and period  without postulating
any value of the distance. The conventions follow Fig.~1. The
simulations use $N=10$ astrometric and $N_{\rm v}=5$ spectroscopic
``observations''. The angular size
of the orbits depends on the actual distance to Sgr A$^*$; for
$D_0=8.5~{\rm kpc}$ the semimajor axis of $1000~{\rm AU}$ corresponds to 
$118~{\rm mas}$.
}

\figcaption[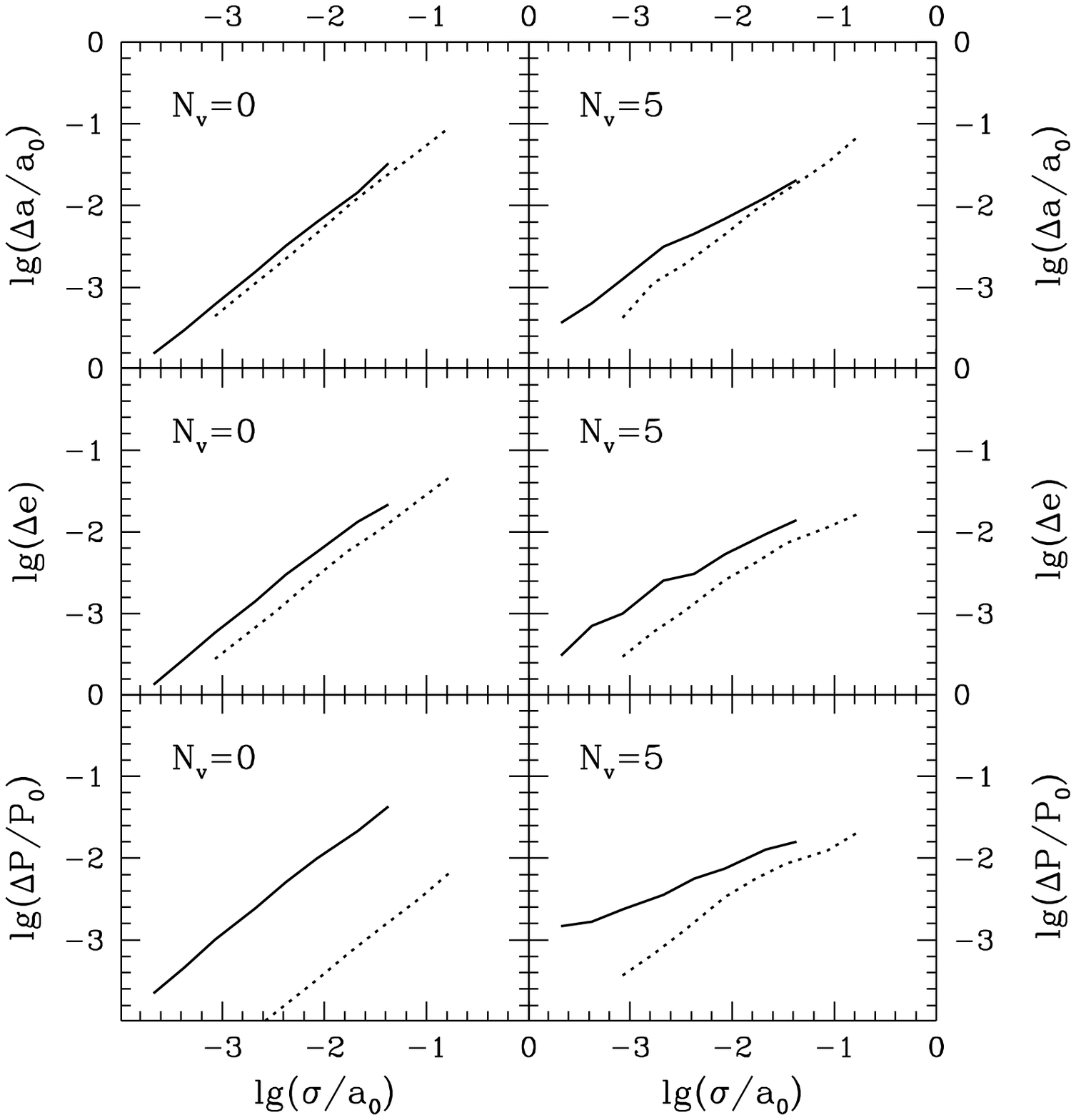]{The scatter in fitted orbital parameters
as a function of relative accuracy in position measurements $\sigma/a_0$.
Each panel shows the scatter in one parameter labeled along the y-axis. 
In the left column the results based on  astrometric ``observations'' alone
($N=10$, $N_{\rm v}=0$) are shown and in the right column the results 
related to combined astrometric and radial velocity ``observations'' 
($N=10$, $N_{\rm v}=5$) are displayed. Simulations use two values 
of the true semimajor axis $a_0=800~{\rm AU}$ (solid lines) 
and $200~{\rm AU}$ (dotted lines). While the
assumed accuracy of the position measurement changes substantially 
we keep the assumed accuracy of radial velocity measurements constant.
}

\figcaption[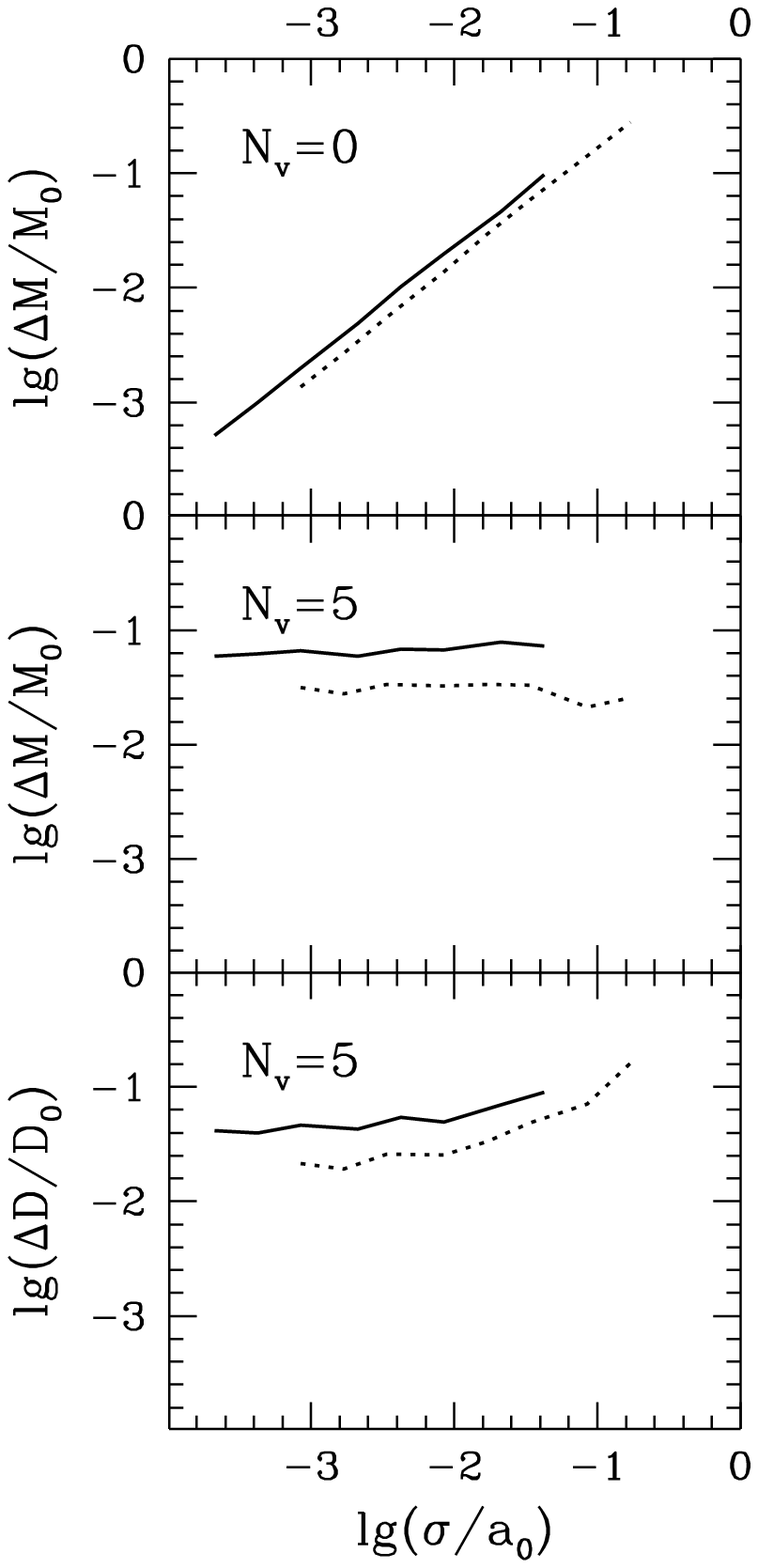]{
The scatter in the estimated mass of the central black hole and in the
estimated distance to the Galaxy Center. The upper panel shows the scatter
in the mass estimate based on astrometry alone, under the assumption that
the distance to the source is known. The middle panel shows the scatter in
mass estimate, which is based on radial velocity and astrometric
measurements and is independent of the distance to the source. The lower
panel shows the scatter in the distance to the source based on the radial
velocity measurements. Conventions follow Fig.~4.
}

\end{document}